\documentclass[aps, prl, twocolumn, secnumarabic, amssymb, superscriptaddress]{revtex4-1}

\usepackage{amsmath}
\usepackage{amssymb}
\usepackage{braket}
\usepackage{mathtools}
\usepackage{bm}
\usepackage {verbatim}

\begin{document}

\title{Attosecond XUV probing of vibronic quantum superpositions in Br$_2^+$}
\author{Yuki Kobayashi}
\email{ykoba@berkeley.edu}
\affiliation{Department of Chemistry, University of California, Berkeley, CA 94720, USA}
\author{Daniel M. Neumark}
\email{dneumark@berkeley.edu}
\affiliation{Department of Chemistry, University of California, Berkeley, CA 94720, USA}
\affiliation{Chemical Sciences Division, Lawrence Berkeley National Laboratory, Berkeley, CA 94720, USA}
\author{Stephen R. Leone}
\email{srl@berkeley.edu}
\affiliation{Department of Chemistry, University of California, Berkeley, CA 94720, USA}
\affiliation{Chemical Sciences Division, Lawrence Berkeley National Laboratory, Berkeley, CA 94720, USA}
\affiliation{Department of Physics, University of California, Berkeley, CA 94720, USA}

\date{\today}

\maketitle

{\bf
Ultrafast laser excitation can create coherent superpositions of electronic states in molecules and trigger ultrafast flow of electron density on few- to sub-femtosecond time scales.
While recent attosecond experiments have addressed real-time observation of these primary photochemical processes, the underlying roles of simultaneous nuclear motions and how they modify and disturb the valence electronic motion remain uncertain.
Here, we investigate coherent electronic-vibrational dynamics induced among multiple vibronic levels of ionic bromine (Br$_2^+$), including both spin-orbit and valence electronic superpositions, using attosecond transient absorption spectroscopy.
Decay, revival, and apparent frequency shifts of electronic coherences are measured via characteristic quantum beats on the Br-$3d$ core-level absorption signals.
Quantum-mechanical simulations attribute the observed electronic decoherence to broadened phase distributions of nuclear wave packets on anharmonic potentials. 
Molecular vibronic structure is further revealed to be imprinted as discrete progressions in electronic beat frequencies.
These results provide a future basis to interpret complex charge-migration dynamics in polyatomic systems.
}

The most elementary step of photochemical reactions is the ultrafast response of valence electrons, which can occur faster than the subsequent rearrangements of molecular geometry \cite{Leone14,Woerner17}.
These pure electronic dynamics are described by coherent superpositions of two or more electronic states, in which the interference between electronic wave functions gives rise to oscillations of total electron density.
Few-electronvolt energy spacings of valence electronic levels dictate that the time scales of electronic motions can be as short as few- to sub-femtosecond. 
Experimental access to coherent electronic dynamics has been enabled by the recent advent of high-harmonic-generation-based ultrafast x-ray sources, which provide unprecedented attosecond time resolution as well as intrinsically broad spectral windows that address atomic core orbitals \cite{Corkum07,Krausz09,Ramasesha16}.
Early attosecond experiments demonstrated that strong-field ionization can prepare coherent superpositions of valence spin-orbit states in rare-gas atoms \cite{Goulielmakis10,Wirth11}.
Coherent electronic motion occurring through chemical bonds, a process known as charge migration, has been reported in a few molecular systems \cite{Calegari14,Kraus15}.

Despite recent success in the characterization of electronic coherences, the roles of simultaneous nuclear motions remain elusive.
In a model case where the nuclear motions are frozen, coherent electronic dynamics continue indefinitely at a fixed period. 
In a real system with moving nuclei, however, molecular vibrations, intramolecular vibrational redistribution, dissociation, and curve crossings occur on femtosecond time scales, affecting the phase and overlap relations between the electronic states, and thus leading to decoherence \cite{Woerner17}.
Furthermore, the nuclear degrees of freedom in molecules add complexity to the energy landscapes of electronic dynamics.
Instantaneous energy spacings of electronic states change in time as molecular vibrations and dissociation occur, which translate to apparent frequency shifts of coherent electronic motions.
The characteristics of bound or dissociative potential surfaces determine the frequency spectra of electronic quantum beats, in an analogous way to discrete vibrational progressions or continuous spectral bands of optical absorption and photoelectron spectra \cite{Herzberg50}.

Experimental studies of coherent electronic-vibrational dynamics are rare, and many of the first attosecond experiments were on atomic systems \cite{Goulielmakis10,Wirth11,Fleischer11,Kobayashi18}.
There are reports on molecules, but significant effects of the molecular vibrations have yet to be discussed \cite{Fleischer11,Kraus13,Timmers19,Ando19,Kobayashi20DBr}.
In optical-domain experiments, electronic superpositions were reported for the Fenna-Matthews-Olsen photosynthetic complex \cite{Engel07}, which generated considerable discussion concerning the role of electronic versus vibrational coherences in large molecules.
In attosecond charge-migration studies, the role of vibrational motions has been considered in the context of decoherence, and theoretical predictions await experimental verifications \cite{Halasz13,Despre15,Vacher17,Arnold17,Despre18,Jia19}.
Revealing the role of nuclear motions in regulating and perturbing electronic coherences is fundamental to attochemistry \cite{Calegari16} and ultimately to designing ultrafast laser control of photochemical dynamics \cite{Remacle06,Lepine14}.

\begin{figure*}[tb]
\includegraphics[scale=1.0]{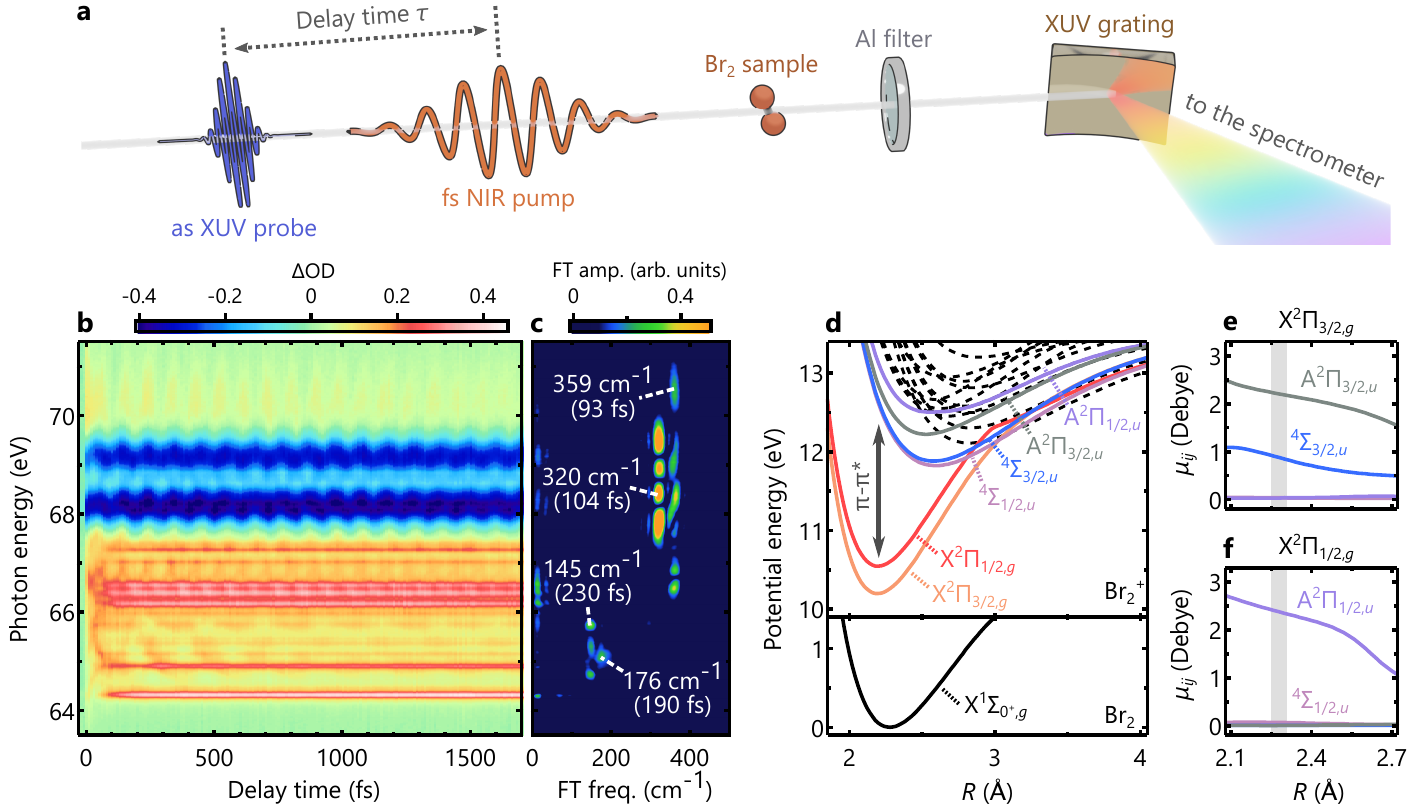}
\caption{\label{fig1}
{\bf Outline of the experiments, long-delay measurements, and electronic structure of bromine.}
{\bf a,} Illustration of the attosecond transient absorption setup.
{\bf b,} Long-delay transient absorption spectra of bromine.
Vibrational dynamics induced among multiple bound ionic states are clearly resolved. 
{\bf c,} Fourier-transform (FT) analysis of the transient absorption spectra.
The FT frequencies and the corresponding vibrational periods are denoted.
{\bf d,} Ab-initio potential energy curves of bromine.
{\bf e,f,} Transition dipole moments from the (e) X$^2\Pi_{3/2,g}$ and (f) X$^2\Pi_{1/2,g}$ states to the excited electronic states.
The gray shades show the Franck-Condon region of the neutral bromine.
}
\end{figure*}

Here, we present an attosecond transient absorption study on coherent electronic-vibrational dynamics of ionic molecular bromine (Br$_2^+$).
This species has several bound electronic states in the valence levels \cite{Coxon71,Harris83,Liang19}, and thus offers an ideal platform to study vibronic superpositions.
Also, no intramolecular vibrational redistribution can occur in diatomic molecules, and the effects of vibrational dephasing and recurrences on electronic superpositions can be interpreted explicitly.

Figure 1a shows the schematic of the experiment (see Methods for the details).
A few-cycle near-infrared (NIR) pulse ($<4$ fs) induces strong-field ionization and excitation of bromine, launching coherent superpositions among its ionic electronic states.
A subsequent attosecond extreme-ultraviolet (XUV) pulse ($<200$ as) arriving after a delay time $\tau$ probes the coherent dynamics via Br-$3d$ core-level absorption at photon energies of $\sim65$ eV.
Pump-on and pump-off XUV spectra are collected at each delay time, and the absorption spectrum is defined by the differential optical density, $\Delta\text{OD}(\omega,t)=-\ln\left[I_{\text{on}}(\omega,t)/I_{\text{off}}(\omega,t)\right]$.
Transitions from core orbitals, whose energies and shapes remain nearly unperturbed throughout the reaction coordinates, serve as sensitive reporters of valence dynamics \cite{Geneaux19}.
Basic experimental conditions required to characterize electronic coherences are (i) the temporal duration of the pump pulse is shorter than the time scales of electronic coherences and (ii) the probe pulse has a sufficiently broad bandwidth to promote the valence-state populations to a common final state. 
The few-cycle NIR pump, attosecond XUV probe configuration of this experiment satisfies these requirements, and electronic coherences manifest as characteristic quantum beats  in the transient absorption signals \cite{Santra11}. 


Figure 1b shows the transient absorption spectra recorded from -30 to 1700 fs at 10-fs intervals.
The vibrational dynamics are characterized in this long-delay measurement, which allows tentative state assignments to be made.
Figure 1c shows the Fourier transform analysis of the transient absorption spectra, and Fig. 1d shows ab-initio potential energy curves of bromine (see Methods for the details). 
Experimentally known vibrational frequencies are noted below, but the calculated potentials are required for the coherent vibronic simulations discussed later.
The spectroscopic parameters of Br$_2^{+}$ are summarized in the Supplementary Table 1.

\begin{figure*}[tb]
\includegraphics[scale=1.0]{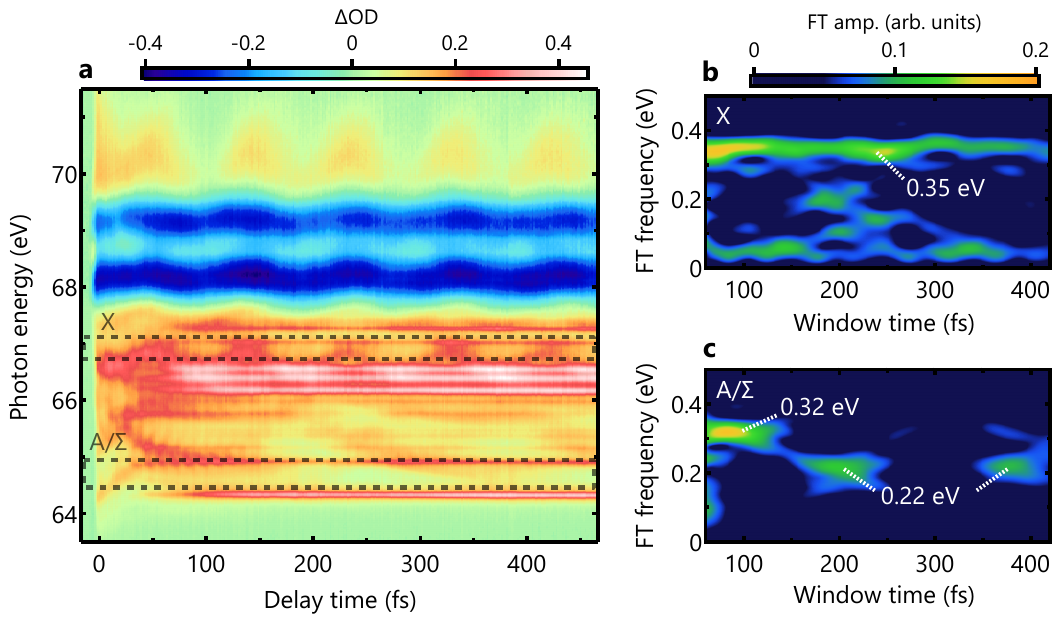}
\caption{\label{fig2}
{\bf Short-time measurements for electronic coherences.}
{\bf a,} Transient absorption spectra of bromine recorded from -30 to 1700 fs at 1.5-fs intervals.
Two spectral regions X (66.7-67.0 eV) and A/$\Sigma$ (64.5-64.9 eV) are marked by dashed boxes.
{\bf b,c} Window Fourier-transformation analysis of the transient absorption spectra for (b) region X and (c) region A/$\Sigma$.
In region X, a constant beat frequency of 0.35 eV is observed, representing electronic coherence in the X$^2\Pi_{g}$ states.
In region A/$\Sigma$, dramatic variations of the beat frequency and amplitude are observed.
These results are indicative of important roles of vibrational motions to maneuver electronic quantum beats.
}
\end{figure*}

The higher frequency features at 320 and 359 cm$^{-1}$ match the harmonic frequency of the neutral X$^1\Sigma_{0^+,g}$ state (323.3 cm$^{-1}$) and those of the ionic X$^2\Pi_{3/2,g}$ and X$^2\Pi_{1/2,g}$ states (364.9 and 361.2 cm$^{-1}$), respectively \cite{Coxon71,Harris83}.
These ground-state vibrational features have already been characterized in a previous study with 40-fs NIR excitation \cite{Hosler13}.
The lower vibrational frequencies at 176 and 145 cm$^{-1}$ are newly observed in the this experiment, and they are assigned to loosely-bound ionic excited states, A$^2\Pi_{u}$ and $^4\Sigma_{u}$ (Fig. 1d).
The experimentally known harmonic frequencies are 205.0, 160.8, 200.6, and 197.8 cm$^{-1}$ for the $^2\Pi_{3/2,u}$, $^2\Pi_{1/2,u}$, $^4\Sigma_{3/2,u}$, and $^4\Sigma_{1/2,u}$ states, respectively \cite{Liang19}.
The deviations between the measured and reference harmonic frequencies implicate a vibrational anharmonicity and a significant manifold of higher vibrational states excited among these electronic states.
Ab-initio simulations of transient absorption spectra could provide unambiguous state assignments, but are not pursued in this study because such simulations for an open-shell halogen are computationally too demanding.

Electronic coherences are characterized in short delay-time measurements recorded from -17 to 466 fs at 1.5-fs intervals (Fig. 2a). 
Note that coherences between the different $g$/$u$ symmetry states, i.e., X$^2\Pi_{g}$ and A$^2\Pi_{u}$/$^4\Sigma_{u}$, cannot be detected in the present probing scheme because the $g\leftrightarrow u$ selection rule prohibits the different $g$/$u$ states to reach a common final state. 
The main focus is placed on electronic quantum beats in two spectral regions marked as X (66.7-67.0 eV) and A/$\Sigma$ (64.5-64.9 eV) (Fig. 2a), wherein the faster (359 cm$^{-1}$) and slower (145 and 176 cm$^{-1}$) vibrational signals are observed, respectively.
Instantaneous beat frequencies are analyzed by taking window Fourier transformation (FT) with a 85-fs super-Gaussian function, and the results are shown in Figs. 2b,c.
The window-FT results around zero delay time are not discussed because of the strong zero-frequency background originating from the sharp rise of the ionic signals (See Supplementary Figure 1 for the full spectrogram).

In region X, a constant beat frequency of 0.35 eV (11.9-fs periodicity) is observed (Fig. 2b). 
This value matches the spin-orbit splitting in the X$^2\Pi_{g}$ states (0.350 eV) \cite{Harris83}, and it can be assigned to electronic coherence between the X$^2\Pi_{3/2,g}$ and X$^2\Pi_{1/2,g}$ states. 
This result reaffirms that strong-field ionization driven by a few-cycle NIR pulse can create coherent superpositions of the spin-orbit doublet states in the ground-state ions, as already demonstrated for several atomic and molecular systems \cite{Goulielmakis10,Fleischer11,Wirth11,Timmers19,Kobayashi20DBr}.

The electronic quantum beats in region A/$\Sigma$ exhibit an entirely different behavior (Fig. 2c).
A quantum beat of 0.32 eV is observed around $100$ fs and it decays in time.
A new beat feature at 0.22 eV emerges transiently around $200$ fs, and the same frequency component revives around $400$ fs.
The shifts in the observed beat frequencies as well as the decay and revival of the quantum beats are particularly important as they are indicative of the unique role of the vibrational motions underlying electronic decoherence and recoherence. 


As an initial guess for the origin of the observed quantum beats in region A/$\Sigma$, an excitation scenario is considered in which coherent wave packets are launched on the A$^2\Pi_{3/2,u}$ and $^4\Sigma_{3/2,u}$ potentials via single-photon excitations from the X$^2\Pi_{3/2,g}$ state (Fig. 1d).
This proposal is based on the following two reasons.
First, the energy spacings between the ground X$^2\Pi_{g}$ states and the excited A$^2\Pi_{u}$ and $^4\Sigma_{u}$ states (1.5-2.3 eV) are within the coverage of the broadband NIR spectrum.
Direct ionization to the ionic excited states is unlikely, as it was not observed in the previous experiment with a narrowband (40-fs) NIR excitation \cite{Hosler13}.
Second, the X$^2\Pi_{g}$ to A$^2\Pi_{u}$/$^4\Sigma_{u}$ one-photon excitations have large transition probabilities, as they correspond to strong $\pi$-$\pi^*$ excitations (Fig. 1d).
The calculated transition dipole moments from the X$^2\Pi_{3/2,g}$ and X$^2\Pi_{1/2,g}$ states are shown in Figs. 1e and f, respectively.
The results reveal that the A$^2\Pi_{3/2,u}$ and $^4\Sigma_{3/2,u}$ states can be accessed simultaneously via optical dipole transitions from the X$^2\Pi_{3/2,g}$ state (Fig. 1e) and predict that there should be no electronic coherence between the A$^2\Pi_{1/2,u}$ and $^4\Sigma_{1/2,u}$ states prepared by this dipole-transition mechanism (Fig. 1f).


\begin{figure*}[tb]
\includegraphics[scale=1.0]{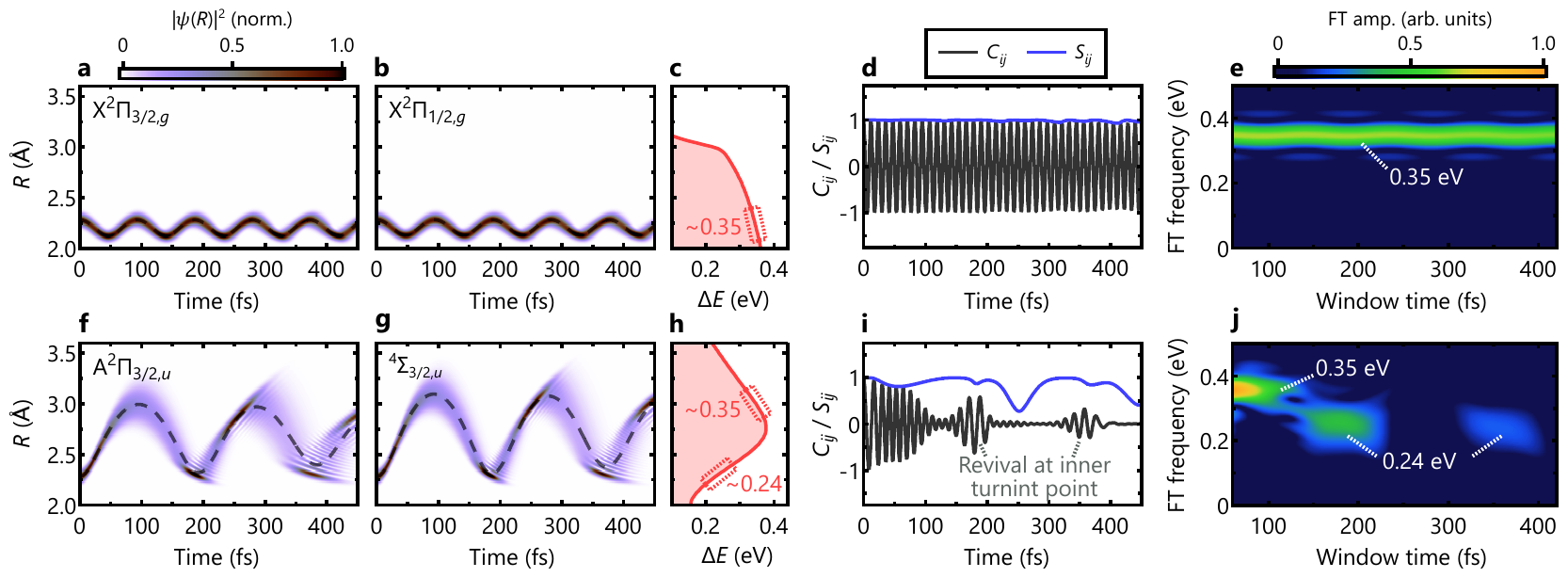}
\caption{\label{fig3}
{\bf Wave-packet simulation of the electronic quantum beats.}
{\bf a,b,} A false-color plot for the simulated probability distributions of the nuclear wave packets on the (a) X$^2\Pi_{3/2,g}$ and (b) X$^2\Pi_{1/2,g}$ states.
Black dashed curves show the center of amplitude of the wave packets.
{\bf c,} Energy spacing ($\Delta E$) between the X$^2\Pi_{3/2,g}$ and X$^2\Pi_{1/2,g}$ states as a function of the internuclear distance.
The dashed box highlights the energy spacing around the region of the vibrational motion. 
{\bf d,} Spatial overlap ($S_{ij}$, blue curve) and electronic coherence ($C_{ij}$, gray curve) between the nuclear wave packets on the X$^2\Pi_{3/2,g}$ and X$^2\Pi_{1/2,g}$ states.
{\bf e,} Window Fourier-transform analysis of the calculated electronic coherence in (d).
The constant 0.35 eV component well reproduces the experimental results.
{\bf f,g,} Simulated probability distributions of the nuclear wave packets on the (f) A$^2\Pi_{3/2,u}$ and (g) $^4\Sigma_{3/2,u}$ states.
{\bf h,} Energy spacing between the A$^2\Pi_{3/2,u}$ and$^4\Sigma_{3/2,u}$ states as a function of the internuclear distance.
The dashed boxes highlight the energy spacings around the inner and outer turning points of the vibrational motions.
{\bf i,} Spatial overlap (blue curve) and electronic coherence (gray curve) between the nuclear wave packets on the A$^2\Pi_{3/2,u}$ and$^4\Sigma_{3/2,u}$ states.
{\bf j,} Window Fourier-transform analysis of the calculated electronic coherence in (i).
The experimentally observed frequency shifting and beat revival are reproduced.
}
\end{figure*}

Coherent wave-packet dynamics of bromine are simulated by numerically solving the time-dependent Schr{\"o}dinger equation (see Methods for the details). 
The states included in the simulations are X$^2\Pi_{3/2,g}$ and X$^2\Pi_{1/2,g}$ for region X, and A$^2\Pi_{3/2,u}$ and $^4\Sigma_{3/2,u}$ for region A/$\Sigma$.
The initial wave packets are prepared by vertical projection of the neutral ground-state wave function.
The nuclear wave packets on two different electronic states, $\psi_i(R,t)$ and $\psi_j(R,t)$, are characterized by the coherence $C_{ij}$ and spatial overlap $S_{ij}$ that are defined as \cite{Vacher17}
\begin{align}
C_{ij}(t)&=\operatorname{Re}\left[\int \psi^*_i(R,t) \psi_j(R,t) dR\right]\\
S_{ij}(t)&=\int \left|\psi^*_i(R,t) \psi_j(R,t)\right| dR.
\end{align}
The real term is taken for the coherence instead of its absolute value in order to visualize the actual beats.
This coherence can be associated with the observed quantum beats assuming that the XUV absorption strengths are largely invariant within the region of the vibrational motions.
Comparison between $C_{ij}$ and $S_{ij}$ helps identify whether decoherence is caused by the loss of spatial overlap ($S_{ij}\approx|C_{ij}|$), or by dephasing ($S_{ij}>|C_{ij}|$) \cite{Fiete03,Vacher17}.

The nuclear wave-packet motions on the X$^2\Pi_{3/2,g}$ and X$^2\Pi_{1/2,g}$ potentials are shown in Figs. 3a and b, respectively.
The energy spacing between the two potentials is relatively flat in the vicinity of the equilibrium distance (Fig. 3c), as it mostly originates from the spin-orbit coupling in the $\pi^*$ orbital.
The X$^2\Pi_{3/2,g}$ and X$^2\Pi_{1/2,g}$ states thus exhibit identical vibrational motions, and the spatial overlap and coherence between the nuclear wave packets are perfectly maintained throughout the simulated time range (Fig. 3d).
The window-FT analysis is performed on the simulated coherence, and the result is shown in Fig. 3e.
The simulation reproduces the observed quantum beat at 0.35 eV, providing clear confirmation of the signal assignment.

The nuclear wave-packet motions on the A$^2\Pi_{3/2,u}$ and $^4\Sigma_{3/2,u}$ potentials are shown in Figs. 3f and g, respectively.
The wave packets sweep across the two loosely bound potentials, with the center of amplitude (dashed curves) reaching $R\approx3.0$ \AA{}.
The wave packets return to the inner turning point at 193 and 186 fs for the A$^2\Pi_{3/2,u}$ and $^4\Sigma_{3/2,u}$ states, respectively, and these calculated periods agree with the experimentally observed period of 190 fs (Fig. 1c).
The coherence and spatial overlap exhibit dramatic variation as the wave packets evolve on the highly anharmonic potentials (Fig. 3i).
The window-FT analysis is shown in Fig. 3j, and the simulated results successfully reproduce the experimental electronic quantum beats including the frequency shifting and the beat revival (Fig. 2c).
The two beat frequencies, 0.35 and 0.24 eV (Fig. 3j), approximately correspond to the energy spacings of the A$^2\Pi_{3/2,u}$ and $^4\Sigma_{3/2,u}$ potentials at the outer-turning points and inner-turning points, respectively (Fig. 3h).
The correspondence between the instantaneous beat frequencies and the inner/outer energy spacings is further verified in a cosine-function fitting analysis (Supplementary Figure S2).

More insights into the coherence dynamics can be obtained by inspecting the wave-packet motions and comparing $S_{ij}$ and $C_{ij}$ (Figs. 3f-i).
The wave packets maintain a high degree of coherence after the excitation until they reach the outer turning points at $\sim100$ fs (Fig. 3i), as they have a high degree of spatial overlap and a well-defined phase relation.
As the wave packets return from the outer turning points, the spatial overlap remains high but the coherence starts to exhibit a gradual decrease (Fig. 3i).
This is because the wave packets are significantly broadened at the outer turning points by the large anharmonicity of the excited potentials, and the developed phase distribution leads to a temporary decoherence. 
When the wave packets return to the inner turning points at $\sim190$ fs (Figs. 3f,g), they are localized at the same initial position of $R=2.3$ \AA{}, a well-defined phase relation is recovered, and the electronic coherence revives (Fig. 3i).
The recovered electronic coherence does not last long, because the spatial overlap sharply decreases when the wave packets leave the inner turning points.
The revival of electronic coherence occurs when the wave packets are localized again at the inner turning point ($\sim380$ fs), while the accrued effects of dephasing between the nuclear wave packets result in a smaller beat amplitude ($\sim50$\% decrease). 


\begin{figure*}[tb]
\includegraphics[scale=1.0]{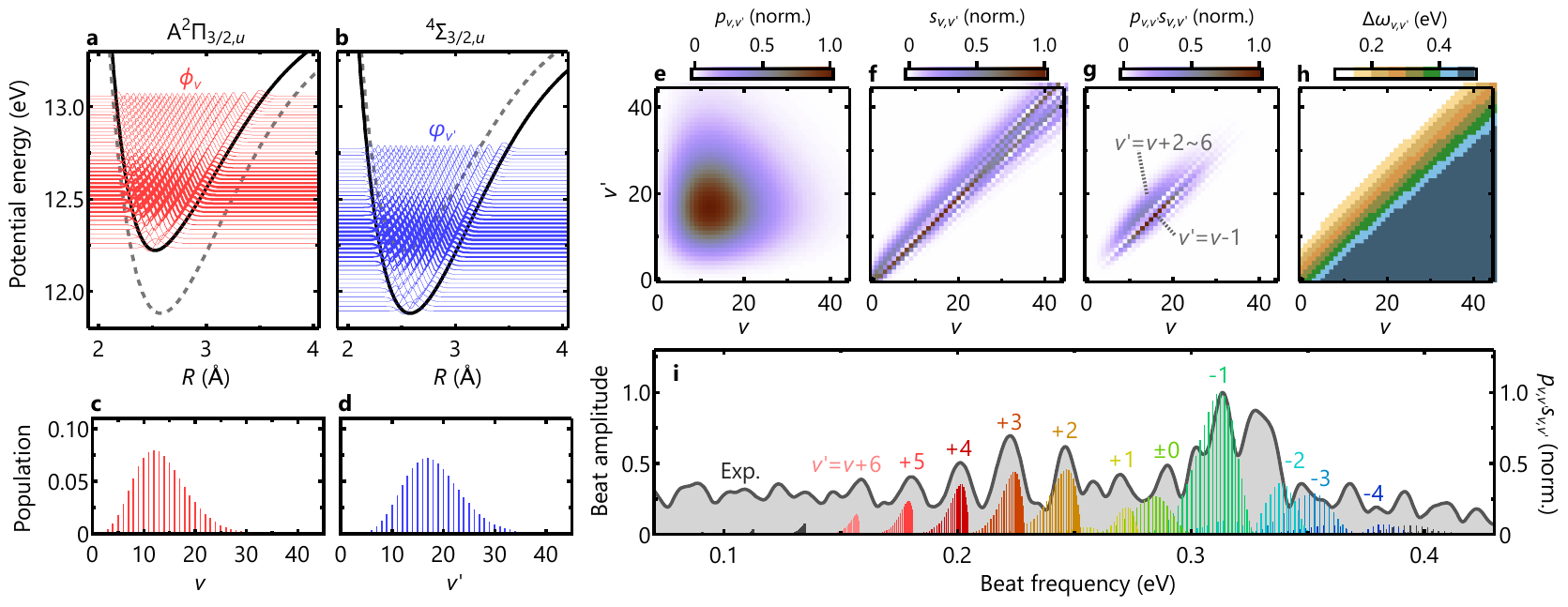}
\caption{\label{fig4}
{\bf Vibronic-structure analysis of the electronic quantum beats.} 
{\bf a,b,} Vibrational eigenfunctions of the (a) A$^2\Pi_{3/2,u}$ and (b) $^4\Sigma_{3/2,u}$ states.
The vibronic levels $v=0$--50 are plotted with their thickness representing the relative populations.
The solid and dashed lines show the potential energy curves of the two states.
{\bf c,d,} Population distributions calculated from the vertical projection of the neutral ground-state wave packet onto the ionic potentials.
{\bf e-h,} Summary of the vibronic state analysis, i.e., (e) cross population term $p_{v,v'}$, (f) cross overlap term $s_{v,v'}$, (g) total products $p_{v,v'}s_{v,v'}$, and (h) the energy spacing $\Delta\omega_{v,v'}$.
{\bf i,} Comparison between the experimental beat frequency (gray area) and the calculated weight of the $v$-$v'$ pairs.
Different coloring is used to distinguish the main $v=v+n$ series ($n=-4\sim+6$).
}
\end{figure*}

So far, the electronic-vibrational dynamics have been viewed as a pair of nuclear wave packets moving on two electronic potentials, and the observed instantaneous beat frequencies are associated with the energy spacings that vary as a function of the internuclear distance.
Alternatively, each wave packet can be viewed as a superposition of multiple vibrational eigenfunctions, and this picture brings to light that electronic quantum beats should consist of a finite number of frequency components that strictly correspond to discrete energy spacings of vibronic levels.

The first step for analyzing the vibronic structure is to decompose the coherence term into pairs of vibrational eigenfunctions $\phi_v$ and $\varphi_{v'}$,
\begin{align}
C_{ij}(t)&=\operatorname{Re}\left[\sum_{v,v'} a_{v}^*(t) b_{v'}(t) \int \phi_v^*(R) \varphi_{v'}(R) dR \right]\\
&=\operatorname{Re}\left[\sum_{v,v'} p_{v,v'} S_{v,v'} e^{i\theta_{v,v'}} e^{-i\Delta\omega_{v,v'}t} \right].
\end{align} 
The vibrational eigenfunctions, $\phi_v$ and $\varphi_{v'}$ for the A$^2\Pi_{3/2,u}$ and $^4\Sigma_{3/2,u}$ states, respectively, as well as the calculated population distributions assuming the vertical projection of the neutral wave function are shown in Figs. 4a-d.
The weight of each $v$-$v'$ vibronic pair is given by a cross population $p_{v,v'}=|a_{v}^*(0) b_{v'}(0)|$ and a cross overlap $S_{v,v'}=\left|\int \phi_v^*(R) \varphi_{v'}(R) dR\right|$.
The phase evolves at a constant rate of $\Delta\omega_{v,v'}$.
The initial phase information is stored in $\theta_{v,v'}$, which is unimportant for the current analysis.

The population term reaches its maximum at $(v,v')=(12,17)$ (Fig. 4e).
Among the numerous $v$-$v'$ pairs populated by the strong-field excitation, the overlap term factors out only those with $v \approx v'$ because of the approximate orthogonality between the vibrational eigenfunctions (Fig. 4f).
The product weight terms $p_{v,v'}s_{v,v'}$ are shown in Fig. 4g.
The $v'=v-1$ series makes the largest contribution, and the $v'=v+2$-$6$ series yield additional contributions.
The energy spacings of the $v$-$v'$ pairs are shown in Fig. 4h.
Within the same $v'=v+n$ series, the energy spacings vary only by a few meV corresponding to the difference in the harmonic frequencies of the two potentials, whereas the neighboring $n$ and $n+1$ series are separated by $\sim 0.023$ eV (or $\sim 180$ cm$^{-1}$), which is equal to the experimentally measured vibrational frequency for the ionic excited states (Fig. 1c).

The calculated weight of the $v$-$v'$ pairs is compared to a frequency spectrum of the experimental quantum beats in region A/$\Sigma$ (Fig. 4i).
In the plot, the calculated energy spacings are shifted by $-0.04$ eV to take into account an overestimate in the adiabatic excitation energies (see Supplementary Table 1).
A good match is seen between the strongest peak at 0.32 eV and the calculated $v'=v-1$ series, and also between the sequence of the peaks from 0.15 to 0.25 eV and the calculated $v'=v+2$--6 series.
Only the $v'=v+n$ series are resolved in the present experiments, but an extended scan of $\sim10$ ps duration, according to simulations, will ultimately reveal individual $v$-$v'$ pairs (see Supplementary Figure 3).
In the conventional Franck-Condon analysis for optical absorption or photoelectron spectra, spatial overlap is taken between vibrational eigenfunctions of the initial and final states of an optical excitation, and a vibrational progression represents simultaneous excitation of electronic and vibrational states in molecules.
Here, the spatial overlap is taken between the nuclear wave packets on two electronic potentials, and the progression signifies the molecular vibronic structure translated as discretized beat frequencies of the coherent electronic motions.

This work establishes a prominent role of vibrational motions in maneuvering the observed electronic quantum beats in molecules.
In the anharmonic potentials of bromine, the broadening of the nuclear wave packets transiently leads to decoherence, and the revival of the electronic quantum beats occur when the wave packets are localized at the inner-turning points.
Potential anharmonicity is a ubiquitous effect and will be a critical factor for designing long-lived electronic coherences in large molecules.
It is also suggested that the electronic coherence between the A$^2\Pi_{3/2,u}$ and $^4\Sigma_{3/2,u}$ states is prepared by a post-ionization, single-photon excitation, which corresponds to the strong $\pi$-$\pi^*$ excitation.
Similar excitation scenarios can potentially be achieved in other $\pi$-conjugated molecules, such as halo-benzene and halo-acetylene, wherein well-defined charge migration can be induced among the electronically excited ionic states.
Lastly, we demonstrate that coherent electronic motion in molecules are subject to additional quantization by the vibrational dynamics.
Knowing the temporal behavior and precise frequencies of coherent electronic motions is fundamental for triggering and capturing charge migration, and this study paves the way toward sub-femtosecond laser engineering of chemical reactivity.

\bibliographystyle{apsrev4-1.bst}
%

\section{Acknowledgments}
This material is based upon a work supported by the National Science Foundation No. CHE-1660417 (Y.K., S.R.L.) and the US Army Research Office No. W911NF-14-1-0383 (Y.K., D.M.N., S.R.L.) and No. W911NF-20-1-0127 (D.M.N.).
Y.K. also acknowledges financial support by the Funai Overseas Scholarship.

\section{Author contributions}
Y.K. performed the experiments and simulations.
D.M.N. and S.R.L. supervised the project.
All authors contributed to the preparation of the manuscript.

\section{Competing interests}
The authors declare no competing interests.

\clearpage
\section{Methods}
{\bf Experiments.} 
Details of the attosecond transient absorption apparatus are described in Refs. \cite{Kobayashi18,Kobayashi20DBr}.
The output of a carrier-envelope phase stable titanium:sapphire laser system (790 nm, 1.8 mJ, 1 kHz) is focused into a neon-filled stretched hollow-core fiber for spectral broadening, and a few-cycle NIR pulse ($<4$ fs, 750 nm, 0.8 mJ) is obtained after phase compensation by a combination of chirped mirrors, fused silica and an ammonium dihydrogen phosphate plate \cite{Timmers17}.
Part of the few-cycle NIR pulse (200 $\mathrm{\mu}$J) is used for high-harmonic generation in argon to produce an attosecond XUV pulse, and the other part (90 $\mathrm{\mu}$J) is used as an ionization pump pulse.
The XUV spectrum is tuned around 65 eV to access the Br-$3d$ core-level absorption edge, and previous streaking measurements recorded 200-as temporal duration for similar XUV spectra \cite{Timmers17}.
Thin aluminum filters (200-nm thickness) are used to remove the residual NIR pulse after high-harmonic generation and transient absorption, while the XUV probe pulse is transmitted.
Pump-on and pump-off spectra ($S_{\text{on}}(\omega)$ and $S_{\text{off}}(\omega)$) are measured at every delay step, from which the differential absorption spectra, $\Delta\text{OD}=-\ln(S_{\text{on}}(\omega)/S_{\text{off}}(\omega))$, are obtained.
The peak field intensity of the NIR pump pulse (90 $\mathrm{\mu}$J) estimated from the focus-size measurement (50 $\mathrm{\mu}$m radius) is $\sim5\times10^{14}$ W/cm$^2$.
The bromine sample (reagent grade) is purchased from Sigma-Aldrich, and is used at room temperature without further purification.

{\bf Electronic-structure calculations of bromine.} 
Electronic structure of bromine is computed by using the spin-orbit generalized multiconfigurational quasidegenerate perturbation theory (SO-GMC-QDPT) implemented in the developer version of GAMESS-US \cite{Schmidt93,Zeng17,Kobayashi19HBr}.
The Douglas-Kroll spin-orbit-adapted model core potential developed by Zeng, Fedorov, and Klobukowski (ZFK-DK3) and basis sets of triple-zeta quality are used in all calculations \cite{Toby131,Toby132,Toby133,Toby134}.
First, a Hartree-Fock calculation is performed for the neutral ground state at $R=2.32$ \AA{}, and the results are used as initial inputs for the subsequent multiconfigurational self-consistent field calculations.
The complete active space consists of 9 electrons in 6 orbitals (i.e., $\sigma$, $\pi$, $\pi^*$, and $\sigma^*$), and all 220 configurations are included in the state-average calculations.
Inclusion of all the configurations is necessary in order to obtain smooth potential results throughout the internuclear distance.
An effective nuclear charge of $Z_{\text{eff}}=40.0$ is used for the bromine atoms in the calculation steps of spin-orbit coupling to accurately reproduce the known spin-orbit splitting of the X$^2\Pi_{g}$ states (2820 cm$^{-1}$ or 0.350 eV) \cite{Harris83}.

{\bf Simulations of coherent nuclear wave-packet dynamics.} 
The coherent nuclear wave packet dynamics of bromine shown in Fig. 3 are simulated by numerically solving the time-dependent Shr\"odinger equation,
\begin{align}
i\frac{\partial}{\partial t}\bm{\Psi}=\bm{H}\bm{\Psi},
\end{align}
where $\bm{\Psi}$ is a column vector containing nuclear wave functions of the valence and core-excited electronic states.
Equations are expressed in atomic units. 
The Hamiltonian of a diatomic system, not including the rotational motion, is given by
\begin{align}
\bm{H}(R,t)=
-\frac{1}{2m}\frac{\partial^2}{\partial R^2}
+\bm{V}(R),
\end{align}
where $m$ is a reduced mass of a bromine molecule and $\bm{V}$ is a diagonal matrix for the potential energies.
The nuclear wave packets are expressed by using the sinc-function discrete variable representation (sinc-DVR) on one-dimensional grids from 1.80 \AA{} to 4.00 \AA{} at the intervals of 0.01 \AA{} \cite{Colbert92}.
The time evolution is computed by using the short-iterative Lanczos method \cite{Leforestier91} at 100-as intervals.
The initial wave packets are defined by vertical projection of the neutral ground-state wave function onto the ionic ground and excited potentials. 


\end{document}